\documentclass[12pt]{article}
\usepackage[totalwidth=460truept,totalheight=625truept]{geometry}
\usepackage{amsfonts,latexsym,amssymb,amsmath,graphicx,accents,slashed,subfigure}
\usepackage{dsfont}
\usepackage[T1]{fontenc}
\usepackage[hidelinks]{hyperref}

\global\arraycolsep=1truept

\numberwithin{equation}{section}

\begin{document}

\phantom{C}

\vskip2truecm

\begin{center}
{\huge \textbf{A New Formulation Of}}

\vskip.4truecm

{\huge \textbf{Lee-Wick Quantum Field Theory}}

\vskip1.5truecm

\textsl{Damiano Anselmi\footnote{%
damiano.anselmi@unipi.it} and Marco Piva\footnote{%
marco.piva@df.unipi.it}}

\vskip .2truecm

\textit{Dipartimento di Fisica \textquotedblleft Enrico
Fermi\textquotedblright , Universit\`{a} di Pisa, }

\textit{Largo B. Pontecorvo 3, 56127 Pisa, Italy}

\textit{and INFN, Sezione di Pisa,}

\textit{Largo B. Pontecorvo 3, 56127 Pisa, Italy}

\vskip2truecm

\textbf{Abstract}
\end{center}

The Lee-Wick models are higher-derivative theories that are claimed to be
unitary thanks to a peculiar cancelation mechanism. In this paper, we
provide a new formulation of the models, to clarify several aspects that
have remained quite mysterious, so far. Specifically, we define them as
nonanalytically Wick rotated Euclidean theories. The complex energy plane is
divided into disconnected regions, which can be related to one another by a
well-defined, albeit nonanalytic procedure. Working in a generic Lorentz
frame, the models are intrinsically equipped with the right recipe to treat
the pinchings of the Lee-Wick poles, with no need of external \textit{ad hoc}
prescriptions. We describe these features in detail by calculating the
one-loop bubble diagram and explaining how the key properties generalize to
more complicated diagrams. The physical results of our formulation are
different from those of the previous ones. The unusual behaviors of the
physical amplitudes lead to interesting phenomenological predictions.

\vfill\eject

\section{Introduction}

\setcounter{equation}{0}

The Lee-Wick (LW) models are special higher-derivative theories, defined in
a peculiar way, which are claimed to lead to a perturbatively unitary $S$
matrix \cite{leewick,LWqed,CLOP}. Precisely, the claim is that they are
equipped with well defined cutting equations \cite{cuttingeq}, such that if
we project the initial and final states onto the subspace $V$ of physical
degrees of freedom, only states belonging to the same space $V$ propagate
through the cuts. Several properties of the models and aspects of their
formulation have not been clarified exhaustively, so far. In this paper we
plan to overcome those problems by reformulating the theories completely.

It is well known that higher-derivative kinetic Lagrangian terms may improve
the ultraviolet behaviors of the Feynman diagrams and may turn
nonrenormalizable theories into renormalizable ones, as in the case of
higher-derivative gravity \cite{stelle}. However, the higher-derivative
corrections, if not treated properly, lead to violations of unitarity or
even mathematical inconsistencies \cite{ugo}. The Lee-Wick idea is
promising, because it claims to reconcile renormalizability and unitarity.

The propagators of the LW\ models contain extra poles, which we call \textit{%
LW poles}, in addition to the poles corresponding to the physical degrees of
freedom and the poles corresponding to the gauge degrees of freedom (such as
the longitudinal and temporal components of the gauge fields and the poles
of the Faddeev-Popov ghosts). The LW poles come in complex conjugate pairs,
which we call \textit{LW pairs}. Cutkosky \textit{et al}. (CLOP) showed in
ref. \cite{CLOP} that the $S$ matrix is not analytic when pairs of LW poles
pinch the integration path on the energy. Analyticity is a property we are
accustomed to, but not a fundamental physical requirement. Nakanishi \cite%
{nakanishi} showed that, if defined in a certain way, the models violate
Lorentz invariance. This problem is more serious, but it can be avoided by
defining the theories in a different way. In ref. \cite{CLOP} it was
proposed to treat the pinching of the LW\ poles by means of a procedure of
limit, which is known as \textit{CLOP prescription}. In simple situations,
the CLOP\ prescription gives an unambiguous, Lorentz invariant and unitary
result, as confirmed by the calculations of Grinstein \textit{et al}. \cite%
{grinstein} in the case of the bubble diagram. However, it is not clear how
to incorporate the CLOP prescription into a Lagrangian and ambiguities are
expected in high-order diagrams \cite{CLOP}. Thus, some key issues
concerning the formulation of the LW models have remained open and are
awaiting to be clarified.

It is more convenient to change approach and define the LW models as
nonanalytically Wick rotated Euclidean higher-derivative theories. First, we
know from ref. \cite{ugo} that a Minkowski formulation of such types of
higher-derivative theories is not viable, since in general it generates
nonlocal, non-Hermitian divergences that cannot be removed by any standard
approach. The Wick rotation from the Euclidean framework is thus expected to
play a crucial role, because it is the only viable path.

However, the Wick rotation of the higher-derivative theories we are
considering turns out to be nonanalytic, because of the LW\ pinching, to the
extent that the complex energy plane is divided into disjoint regions of
analyticity. The Lorentz violation is avoided by working in a generic
Lorentz frame, with generic external momenta, deforming the integration
domain on the loop space momenta in a suitable way and then analytically
continuing in each region separately. We show that, if we do so, the models
are intrinsically equipped with all that is necessary to define them
properly. In particular, there is no need of the CLOP prescription, or any
other prescription to handle the pinching of the LW poles. Actually, the
CLOP prescription should be dropped, because it leads to ambiguities, even
in a simple case such as the bubble diagram with different physical masses.

The behaviors of the amplitudes show some unexpected features, which lead to
interesting phenomenological predictions. In particular, the violation of
analyticity is quite apparent, when the amplitude is plotted. If ever
observed, this behavior could be the quickest way to determine the
experimental value of the energy scale $M$ associated with the
higher-derivative terms, which is the key physical constant of the LW models.

Indeed, the Lee-Wick models have been also studied for their possible
physical applications, which include QED \cite{LWqed}, the standard model 
\cite{LWstandardM}, grand unified theories \cite{LWunification} and quantum
gravity \cite{LWgrav}.

The paper is organized as follows. In section \ref{LWE}, we outline the
formulation of the LW models as nonanalytically Wick rotated Euclidean
theories. In section \ref{LWpin} we study the LW pinching in detail, in the
case of the bubble diagram. In particular, we show how Lorentz invariance is
recovered in each region of the complex energy plane. In section \ref%
{LWaround}, we describe the calculations of the physical amplitudes in a
neighborhood of the LW\ pinching and show that the CLOP\ and similar
prescriptions are ambiguous and not consistent with our approach. In section %
\ref{bubblecomplete} we evaluate the bubble diagram in the new formulation
and show that the physical results are in general different from those that
follow from the CLOP\ and other prescriptions. We also comment on the
phenomenological relevance of the results. In section \ref{morecompl} we
explain why the basic properties of our formulation generalize to more
complicated diagrams.

\section{Lee-Wick models as Wick rotated Euclidean theories}

\setcounter{equation}{0}

\label{LWE}

In this section we outline the new formulation of the LW\ models. We begin
by describing the class of higher-derivative theories that we are
considering. The higher-derivative Lagrangian terms are multiplied by
inverse powers of certain mass scales, which we call \textit{LW scales}. For
simplicity, we can assume that there is just one LW scale, which we denote
by $M$, since the generalization to many LW scales is straightforward.

When $M$ tends to infinity, the propagators must tend to the ones of
ordinary unitary theories. Moreover, the extra poles that are present when $%
M<\infty $ must come in complex conjugate pairs and satisfy Re$%
[p^{2}]\geqslant 0$, Im$[p^{2}]\neq 0$.

A typical propagator of momentum $p$ is equal to the standard propagator
times a real function of $p^{2}$ that has no poles on the real axis. For
concreteness, we take 
\begin{equation}
iD(p^{2},m^{2},\epsilon )=\frac{iM^{4}}{(p^{2}-m^{2}+i\epsilon
)((p^{2})^{2}+M^{4})}.  \label{propa}
\end{equation}%
More general propagators can be considered. In particular renormalization
may lead to\ structures such as%
\begin{equation*}
\frac{iM^{4}}{(p^{2}-m^{2}+i\epsilon )((p^{2}-\mu ^{2})^{2}+M^{4})}.
\end{equation*}%
However, the key features are already encoded in (\ref{propa}) and the
extension does not change the sense of our investigation.

\begin{figure}[t]
\begin{center}
\includegraphics[width=9cm]{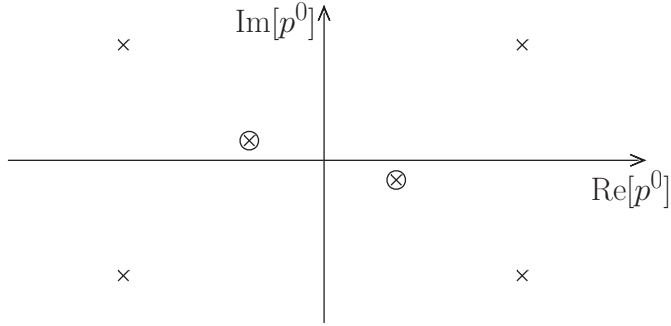}
\end{center}
\caption{Poles of the propagator}
\label{lw}
\end{figure}

The poles of (\ref{propa}) are%
\begin{equation}
p^{0}=\pm \omega _{m}(\mathbf{p})\mp i\epsilon ,\qquad p^{0}=\pm \Omega _{M}(%
\mathbf{p}),\qquad p^{0}=\pm \Omega _{M}^{\ast }(\mathbf{p}),  \label{pol0}
\end{equation}%
where $\omega _{m}(\mathbf{p})=\sqrt{\mathbf{p}^{2}+m^{2}}$ and $\Omega _{M}(%
\mathbf{p})=\sqrt{\mathbf{p}^{2}+iM^{2}}$. Their locations are shown in fig. %
\ref{lw}, where the LW poles are denoted by means of an $\times $, while the
standard poles are denoted by a circled $\times $.

We can integrate $p^{0}$ along the real axis or along the imaginary axis.
The first choice defines the Minkowski theory, the second choice defines the
Euclidean theory. The two give different results, because, even if the
integration path at infinity does not contribute, some poles are located in
the first and third quadrants of the complex plane. In ref. \cite{ugo} it
was shown that in general the Minkowski theories of this type are
inconsistent, because they are plagued with nonlocal, non-Hermitian
divergences that cannot be subtracted away without destroying the basic
properties of the theory. The bubble diagram in four dimensions is one of
the few convergent exceptions, but it becomes nonlocally divergent as soon
as nontrivial numerators are carried by the vertices, which happens for
example in higher-derivative gravity. This fact forces us to proceed with
the Euclidean theory.

Usually, the Wick rotation is an analytic operation everywhere, but in the
Lee-Wick\ models it is analytic only in a region of the complex energy
plane, the one that contains the imaginary axis. We call it \textit{main
region} and denote it by $\mathcal{A}_{0}$. The complex plane turns out to
be divided into several disconnected regions $\mathcal{A}_{i}$, which can be
reached from the main region in a nonanalytic way. The regions $\mathcal{A}%
_{i}$ are called \textit{analytic regions}.

In the light of this fact, the calculation of the correlation functions
proceeds as follows. The loop integrals are evaluated at generic (possibly
complex) external momenta, in each analytic region $\mathcal{A}_{i}$ of the
complex plane. For a reason that we will explain, we anticipate that it is
also necessary to work in a sufficiently generic Lorentz frame, because
special Lorentz frames may squeeze entire regions to lines and make the
calculation ill defined. The $\mathcal{A}_{i}$ subdomain where the
calculation is done is denoted by $\mathcal{O}_{i}$ and has to satisfy
suitable properties. For example, it must contain an accumulation point.

In general, Lorentz invariance and analyticity are lost in the intermediate
steps, in all the regions $\mathcal{A}_{i}$ apart from the main one. They
are recovered by deforming the integration domain on the loop space momenta
in a nontrivial way. After the evaluation, the amplitude is analytically
continued from $\mathcal{O}_{i}$ to the rest of the region $\mathcal{A}_{i}$%
. This procedure gives the amplitude of the LW\ model, region by region.
Since it is not possible to relate the regions analytically, the Wick
rotation is nonanalytic. Yet, the regions are related by a well-defined
nonanalytic procedure, which we describe in the next sections.

We may condense their articulated definition by saying that \textit{the LW
models are nonanalytically Wick rotated Euclidean higher-derivative theories
of a special class}.

\begin{figure}[t]
\begin{center}
\includegraphics[width=8cm]{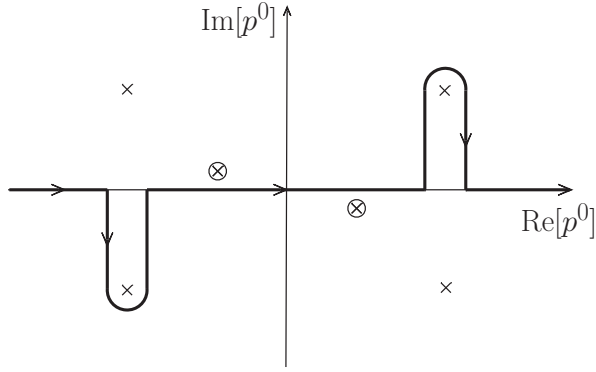}
\end{center}
\caption{The Lee-Wick integration path}
\label{wick}
\end{figure}

\begin{figure}[b]
\begin{center}
\includegraphics[width=4truecm]{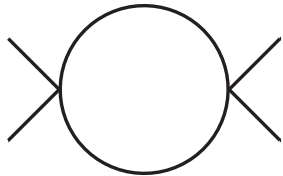}
\end{center}
\caption{Bubble diagram}
\label{bubble}
\end{figure}

\begin{figure}[t]
\begin{center}
\includegraphics[width=11cm]{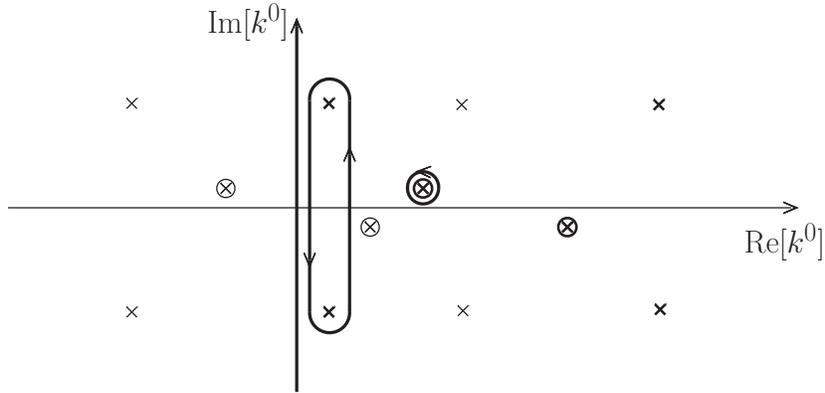}
\end{center}
\caption{Euclidean integration path of the bubble diagram}
\label{deform}
\end{figure}

\begin{figure}[t]
\begin{center}
\includegraphics[width=11cm]{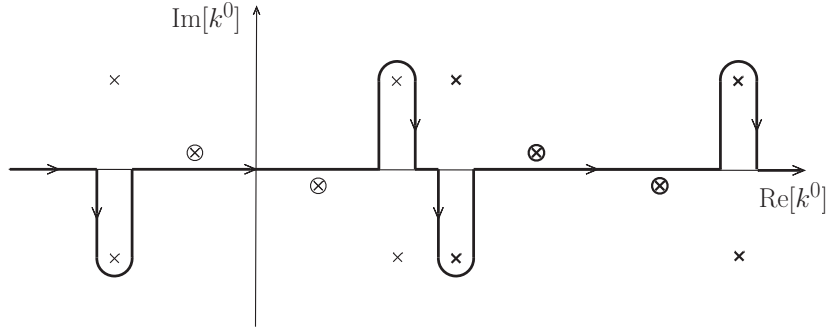}
\end{center}
\caption{Integration path of the bubble diagram after the Wick rotation}
\label{WickBub2}
\end{figure}

Consider the propagator (\ref{propa}) and its poles (\ref{pol0}). When the
imaginary axis is rotated to the real one, we get the integration path shown
in fig. \ref{wick}. The Wick rotation is less trivial when performed in
Feynman diagrams. To be explicit, consider the bubble diagram (fig. \ref%
{bubble}). It has two propagators, so the number of poles doubles. If one
propagator has momentum $k$ and the other propagator has momentum $k-p$, in $%
D$ spacetime dimensions we have a loop integral proportional to 
\begin{equation}
\mathcal{J}(p)=\int \frac{\mathrm{d}^{D}k}{(2\pi )^{D}}D(k^{2},m_{1}^{2},%
\epsilon _{1})D((k-p)^{2},m_{2}^{2},\epsilon _{2}),  \label{bubd}
\end{equation}%
the associated amplitude being $\mathcal{M}(p)=-i\lambda ^{2}\mathcal{J}%
(p)/2 $, where $\lambda $ is the coupling and $1/2$ is the combinatorial
factor. When we vary the external momentum $p$, the poles of the first
propagator are fixed [given by formula (\ref{pol0}) with $p\rightarrow k$, $%
m\rightarrow m_{1}$], while those of the second propagator, which are%
\begin{equation}
k^{0}=p^{0}\pm \omega _{m_{2}}(\mathbf{k-p})\mp i\epsilon ,\qquad
k^{0}=p^{0}\pm \Omega _{M}(\mathbf{k-p}),\qquad k^{0}=p^{0}\pm \Omega
_{M}^{\ast }(\mathbf{k-p}),  \label{pol1}
\end{equation}%
move on the complex $k^{0}$ plane. With respect to the fixed poles, this
sextet of poles is translated by $p^{0}$ and deformed by $\mathbf{p}$. At
some point, the translation makes some poles cross the imaginary axis, which
is the integration path. To preserve analyticity, the integration path must
be deformed so that the crossing does not actually take place. Equivalently,
we can keep the main integration path on the imaginary axis and add
integration contours around the poles that cross the imaginary axis. In the
end, we obtain a path like the one of fig. \ref{deform}, where the thick
poles are the moving ones. Finally, when we make the Wick rotation to the
real axis, we obtain an integration path like the one shown in fig. \ref%
{WickBub} or, depending on $p$, fig. \ref{WickBub2}. In these pictures we
have assumed for simplicity that the external space momentum $\mathbf{p}$
vanishes. The integration paths obtained from the Wick rotation agree with
those prescribed by Lee and Wick. The general rule, valid for arbitrary
Feynman diagrams, is that the left LW\ pair of a propagator is always above
the integration path, while the right LW pair is always below.

\begin{figure}[b]
\begin{center}
\includegraphics[width=10cm]{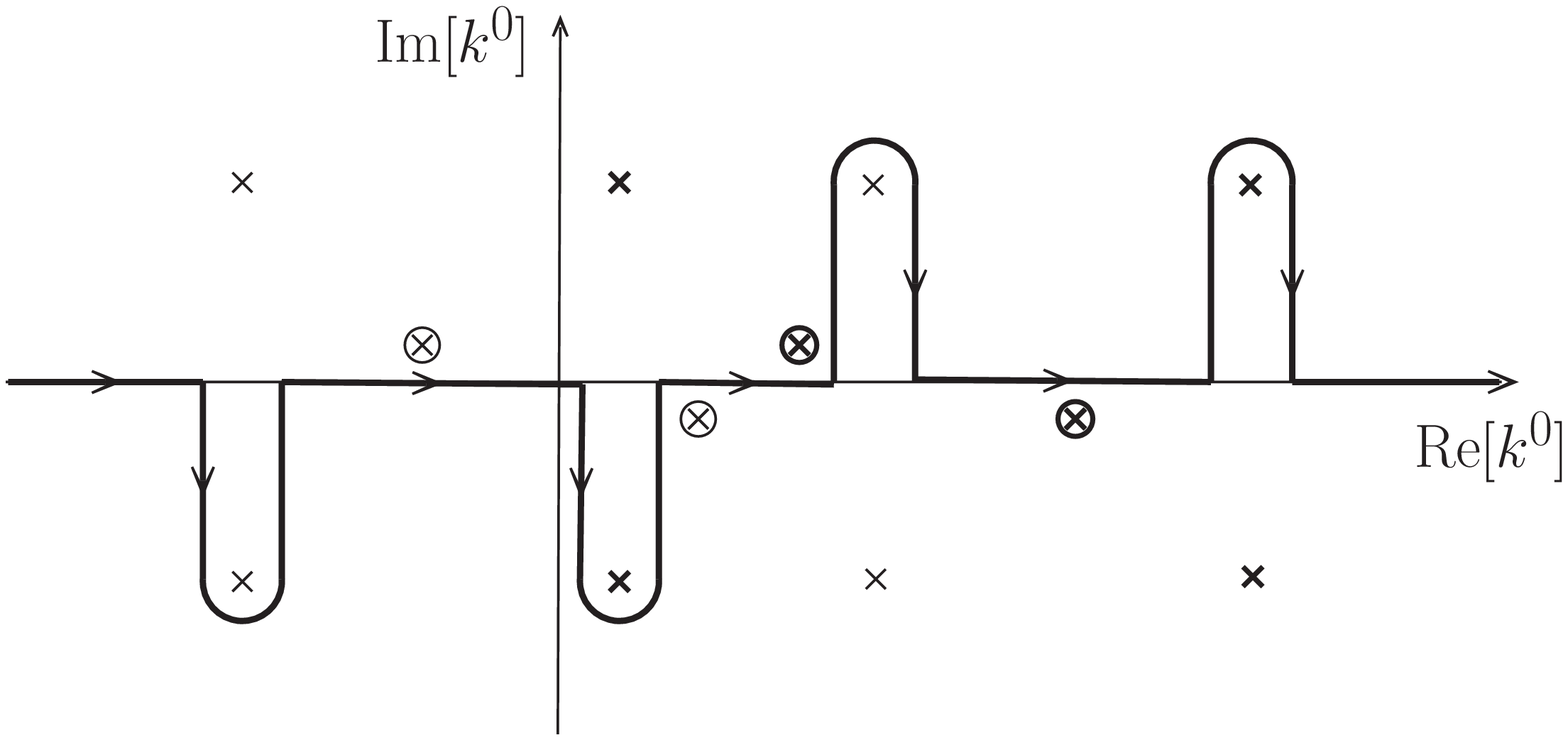}
\end{center}
\caption{Integration path of the bubble diagram after the Wick rotation}
\label{WickBub}
\end{figure}

When the right (respectively, left) LW\ pair of the propagator $%
D(k^{2},m_{1}^{2},\epsilon _{1})$ hits the left (right) LW\ pair of $%
D((k-p)^{2},m_{2}^{2},\epsilon _{2})$, the integration path gets pinched. We
call this occurrence \textit{LW\ pinching}.

The integration paths before and after the LW\ pinching are illustrated in
figs. \ref{WickBub} and \ref{WickBub2}. When we perform the Wick rotation,
the analytic continuation is straightforward in the situation of fig. \ref%
{WickBub}, but we find an unexpected behavior in the situation of fig. \ref%
{WickBub2}. The two situations correspond to disjoint regions $\mathcal{A}%
_{1}$ and $\mathcal{A}_{2}$ of the complex $p^{0}$ plane. Each region $%
\mathcal{A}_{i}$ must be studied separately and gives a complex function $%
\mathcal{J}_{i}(p)$. The complex functions $\mathcal{J}_{1}(p)$ and $%
\mathcal{J}_{2}(p)$ are not related to each other by an analytic
continuation. However they are still related in a well defined, nonanalytic
way.

We show that, with these caveats, the procedure to handle the LW pinching is
intrinsic to our definition of the theory, pretty much like the $i\epsilon $
prescription is intrinsic to the definition of a theory as a Wick rotated
Euclidean one. Moreover, it is consistent with perturbative unitarity.

The LW\ pinching motivated some authors to propose \textit{ad hoc}
prescriptions to handle it. The CLOP prescription \cite{CLOP}, for example,
amounts to deform the scale $M$ in one of the propagators of the integral (%
\ref{bubd}) to a different value $M^{\prime }$. Under certain conditions,
the pinching is absent for $M^{\prime }\neq M$, the regions we mentioned
above are analytically connected and the Wick rotation is analytic
everywhere. After the calculation of the amplitude, the deformed scale $%
M^{\prime }$ is sent to $M$. This operation cuts the complex plane into
disconnected regions.

The CLOP\ prescription is not sufficient to deal with the LW pinching in all
the diagrams, because higher-order diagrams are expected to be ambiguous 
\cite{CLOP}. Moreover, it appears to be artificial. For example, there is no
obvious way to incorporate it into the Lagrangian or the Feynman rules. In
this paper, we also show that the CLOP prescription leads to physical
predictions that differ from the ones we obtain and are ambiguous even in
the case of the bubble diagram with $m_{1}\neq m_{2}$. We also show that, if
we strictly apply the rules that follow from the formulation of this paper,
it is possible to retrieve the correct result even starting from $M^{\prime
}\neq M$ and letting $M^{\prime }$ tend to $M$ at the end. Then, however,
the CLOP prescription becomes redundant.

In section \ref{morecompl} we explain how the results of this section extend
from the bubble diagram to more complicated diagrams.

To summarize, we show that the nonanalytically Wick rotated theory is well
defined and intrinsically equipped with the procedure that allows us to
handle the LW pinching. Instead, the prescriptions that can be found in the
literature are ambiguous or redundant and give predictions that may be in
contradiction with ours.

\section{LW pinching}

\setcounter{equation}{0}

\label{LWpin}

In this section we describe the LW pinching in the case of the bubble
diagram (fig. \ref{bubble}), that is to say the loop integral (\ref{bubd}).
First, we integrate on the loop energy $k^{0}$ by means of the residue
theorem. This operation leaves us with the integral on the loop space
momentum $\mathbf{k}$. Orienting the external space momentum $\mathbf{p}$
along the vertical line, the integral on the azimuth is trivial, so we
remain with the integral on $k_{s}\equiv |\mathbf{k}|$ from $0$ to $\infty $
and the integral on $u\equiv \cos \theta $ from $-1$ to $1$, where $\theta $
is the zenith angle. To illustrate the problematics involved in the LW\
pinching exhaustively, we consider two cases. In the first case we work at $%
\mathbf{p}=0$, in the second case we work at $\mathbf{p}\neq 0$. Lorentz
invariance suggests that there should be no big difference between the two
situations. It turns out that it is not so, because the method of
calculation we are using is not manifestly Lorentz invariant. The
calculation at $\mathbf{p}=0$ misses some crucial points, which are visible
only at $\mathbf{p}\neq 0$.

\subsection{LW\ pinching at zero external space momentum}

The LW pinching may involve pairs of LW poles (in which case it is called 
\textit{pure} LW pinching) or one LW pole and a standard pole (in which case
it is called \textit{mixed} LW pinching). For the moment, we focus on the
pure LW pinching, because at one loop the mixed one cannot occur for real
external momenta.

\begin{figure}[t]
\begin{center}
\includegraphics[width=10truecm]{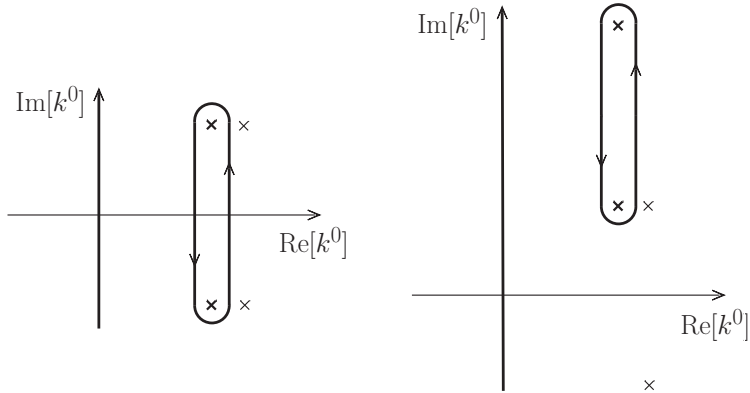}
\end{center}
\caption{Lee-Wick pinching}
\label{pinching2}
\end{figure}

There are two basic cases of pure LW\ pinching, shown in fig. \ref{pinching2}%
. The first case involves the right LW pair of the first propagator and the
left LW\ pair of the second propagator. The second case involves the
upper-right LW pole of the first propagator and the bottom-left LW pole of
the second propagator. The other LW pinchings are the complex conjugates of
the ones just described and their reflections with respect to the imaginary
axis.

At $\mathbf{p}=0$, there is no $u$ dependence, so the $u$ integral is
trivial, the only nontrivial integration variable being $k_{s}$. The poles
relevant to the top pinching occurring in the left figure \ref{pinching2} are%
\begin{equation}
\frac{1}{k^{0}-p^{0}+\Omega _{M}^{\ast }(\mathbf{k})}\frac{1}{k^{0}-\Omega
_{M}(\mathbf{k})},  \label{polpot}
\end{equation}%
while those relevant to the bottom pinching give the complex conjugate of
this expression. The pinching occurs when $k^{0}$ is such that the locations
of the two poles coincide, which gives the pinching equation%
\begin{equation}
p^{0}=\sqrt{k_{s}^{2}+iM^{2}}+\sqrt{k_{s}^{2}-iM^{2}},  \label{peq0}
\end{equation}%
solved by 
\begin{equation}
k_{s}^{2}=\frac{(p^{0})^{4}-4M^{4}}{4(p^{0})^{2}}.  \label{poles0}
\end{equation}

The poles relevant to the pinching occurring in the right figure \ref%
{pinching2} are 
\begin{equation*}
\qquad \frac{1}{k^{0}-p^{0}+\Omega _{M}(\mathbf{k})}\frac{1}{k^{0}-\Omega
_{M}(\mathbf{k})}.
\end{equation*}%
They give the pinching equations%
\begin{equation}
\qquad p^{0}=2\sqrt{k_{s}^{2}+iM^{2}},  \label{peq}
\end{equation}%
which are solved by 
\begin{equation}
k_{s}^{2}=\frac{(p^{0})^{2}}{4}-iM^{2}.  \label{poles}
\end{equation}

We denote the $k_{s}$ integration path by $\Gamma _{k}$. By default, we
expect it to be the positive real axis, but in a moment we will discover
that we must deform it to include complex values.

\begin{figure}[t]
\begin{center}
\includegraphics[width=9truecm]{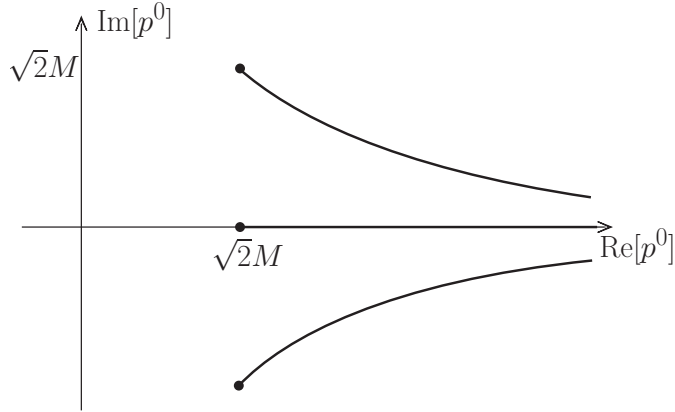}
\end{center}
\caption{Branch cuts due to the Lee-Wick pinching at $\mathbf{p}= 0$}
\label{compl}
\end{figure}

When $k_{s}$ is real and positive, the solution of (\ref{peq0}) exists for $%
p^{2}$ real and larger than $2M^{2}$, while the solution of (\ref{peq})
exists when $p^{2}-4iM^{2}$ is real and larger than zero. Thus, the integral
in $\mathcal{J}(p)$ has the LW\ branch cuts shown in fig. \ref{compl} and
symmetric ones with respect to the imaginary axis. The middle branch point
corresponds to the \textit{LW threshold} $p^{2}=2M^{2}$, while the other two
branch points correspond to the LW thresholds $p^{2}=4iM^{2}$ and $%
p^{2}=-4iM^{2}$. We have not shown the branch cuts associated with the
standard pinching and the mixed LW pinching. When we vary $p^{0}$ across a
branch cut of fig. \ref{compl}, a pole $\nu $ of the $k_{s}$ integrand
crosses the $k_{s}$ integration path $\Gamma _{k}$ (which means that the
imaginary part of the pole becomes zero, while its real part stays
positive), so the function $\mathcal{J}(p)$ is not analytic in that point.

For example, the right-hand side of (\ref{poles0}) has vanishing imaginary
part and positive real part for $x\geqslant \sqrt{2}M$, $y=0$, where $%
x\equiv \mathrm{Re}[p^{0}]$, $y\equiv \mathrm{Im}[p^{0}]$. This gives the
middle branch cut of fig. \ref{compl}, which starts from $p^{0}=\sqrt{2}M$.
A mirror branch cut is obtained by reflecting with respect to the imaginary
axis.

On the other hand, the right-hand side of (\ref{poles}) has vanishing
imaginary part and positive real part when%
\begin{equation}
xy=2M^{2},\qquad x^{2}\geqslant y^{2}.  \label{complexcut}
\end{equation}%
This gives the branch cut shown in the first quadrant of fig. \ref{compl},
which starts from $p^{0}=\sqrt{2}M(1+i)$, and a symmetric branch cut in the
third quadrant. The complex conjugate LW pinching gives the branch cut shown
in the fourth quadrant of fig. \ref{compl}, with branch point $p^{0}=\sqrt{2}%
M(1-i)$, and a symmetric branch cut in the second quadrant.

So far, we have described what happens when $\Gamma _{k}$ is not deformed.
We have seen that in that case certain poles $\nu $ of the integrand cross $%
\Gamma _{k}$ when $p^{0}$ crosses the cuts of fig. \ref{compl}. There, the
function $\mathcal{J}(p)$ is not analytic. This is what we naturally obtain,
for example, if we make the integration numerically, since a generic program
of numerical integration does not know how to analytically deform the
integration paths.

If we want to turn $\mathcal{J}(p)$ into a function that is analytic in a
subdomain $\mathcal{O}$ that intersects the branch cuts of fig. \ref{compl},
we have to move those branch cuts away from $\mathcal{O}$. This is done by
deforming $\Gamma _{k}$ when the poles $\nu $ approach it, so as to prevent $%
\nu $ from crossing $\Gamma _{k}$ in $\mathcal{O}$, and make the crossing
occur at different values of $p^{0}$. Or, we can keep the integration path $%
\Gamma _{k}$ rigid, but add or subtract (depending on the direction of
motion of $\nu $) the residues of the moving poles $\nu $ after the
crossing. For example, in the equal mass case $m_{1}=m_{2}=m$, it is easy to
check that analyticity on the real axis above the LW threshold $p^{2}=2M^{2}$
is effectively restored by the replacement%
\begin{equation*}
\mathcal{J}(p)\rightarrow \mathcal{J}(p)-\frac{1}{16\pi }\frac{M^{4}}{%
m^{4}+M^{4}}\sqrt{1-\frac{4M^{4}}{(p^{2})^{2}}}\theta _{-}(p^{2}-2M^{2}),
\end{equation*}%
when $p^{0}$ crosses the real axis above $\sqrt{2}M$ from the upper half
plane in the first quadrant (or below $-\sqrt{2}M$ from the lower half plane
in the third quadrant), where $\theta _{-}(x)=1$ for $\mathrm{Re}[x]>0$, $%
\mathrm{Im}[x]<0$ and $\theta _{-}(x)=0$ in all other cases. In both sides
of this replacement the integration path $\Gamma _{k}$ is the positive real
axis.

\begin{figure}[t]
\begin{center}
\includegraphics[width=7truecm]{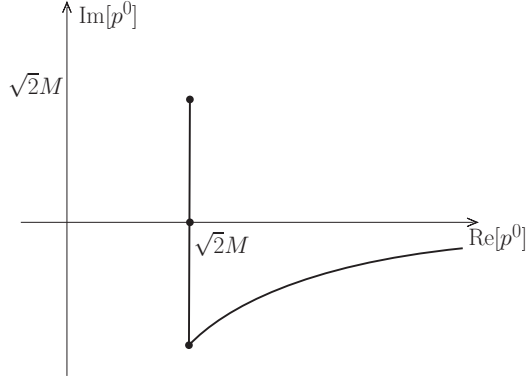}
\end{center}
\caption{Analytic deformation of the branch cuts}
\label{compl4}
\end{figure}

Deforming the cuts with this procedure, we may obtain, for example, fig. \ref%
{compl4}. Now the amplitude $\mathcal{M}(p)=-i\lambda ^{2}\mathcal{J}(p)/2$
is mathematically well defined on the real axis, but it has a nontrivial
imaginary part for $p^{0}$ real and such that $(p^{0})^{2}>2M^{2}$, which
violates unitarity. To preserve unitarity, we must keep the branch cuts
symmetric with respect to the real axis. At $\mathbf{p}=0$ this implies that
a branch cut is necessarily on the real axis, which makes the amplitude ill
defined there.

\subsection{LW\ pinching at nonzero external space momentum}

\begin{figure}[t]
\begin{center}
\includegraphics[width=16truecm]{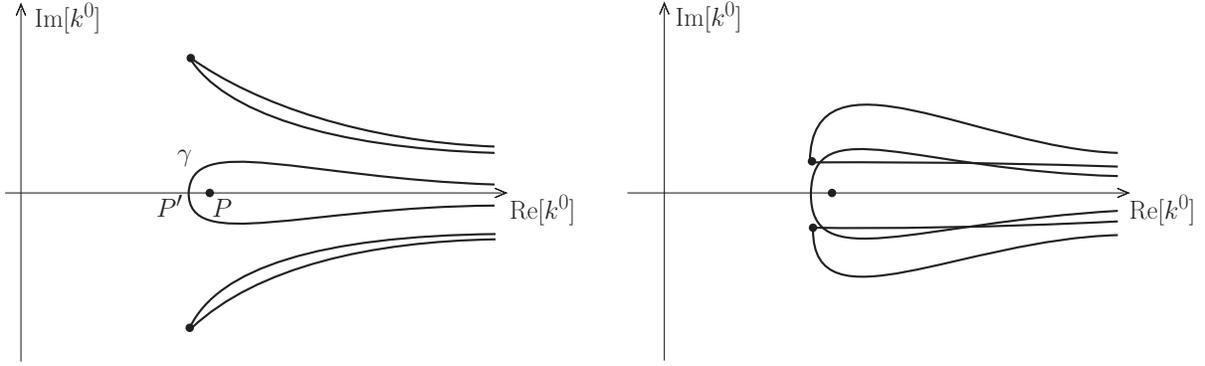}
\end{center}
\caption{Solutions of the Lee-Wick pinching conditions at $\mathbf{p}\neq 0$}
\label{compl2}
\end{figure}

At $\mathbf{p}\neq 0$ several interesting phenomena occur, which eventually
lead to the solution of the problem of properly handling the LW pinching.
The pinching equations (\ref{peq0}) and (\ref{peq}) become%
\begin{equation}
p^{0}=\sqrt{\mathbf{k}^{2}+iM^{2}}+\sqrt{(\mathbf{k-p})^{2}-iM^{2}},\qquad
p^{0}=\sqrt{\mathbf{k}^{2}+iM^{2}}+\sqrt{(\mathbf{k-p})^{2}+iM^{2}},
\label{pinchk}
\end{equation}%
respectively, plus their complex conjugates. Keeping $\mathbf{p}$ fixed, the
solutions fill extended surfaces, shown in fig. \ref{compl2}. The first
picture is obtained for smaller values of $|\mathbf{p|}$, the second picture
for larger values.

Since the right-hand sides of (\ref{pinchk}) now depend on two parameters, $%
k_{s}$ and $u$, the lines of fig. \ref{compl} have enlarged into regions $%
\mathcal{\tilde{A}}_{i}$ of nonvanishing measure. Let $\mathcal{\tilde{A}}%
_{0}$ denote the region that contains the imaginary axis, which we call 
\textit{main region}, and $\mathcal{\tilde{A}}_{P}$ the one that contains
the point $P$, located at $p^{0}=\sqrt{2M^{2}+\mathbf{p}^{2}}\equiv E_{P}$.
Such a point corresponds to the LW threshold $p^{2}=2M^{2}$. Finally, we
call $\mathcal{\tilde{A}}_{P}^{\prime }$ the region symmetric to $\mathcal{%
\tilde{A}}_{P}$ with respect to the imaginary axis. The regions $\mathcal{%
\tilde{A}}_{i}$ other than $\mathcal{\tilde{A}}_{0}$ collect the values of $%
p^{0}$ that satisfy the equations (\ref{pinchk}) for real $\mathbf{k}$.
There, $\mathcal{J}(p)$ gives nonanalytic, Lorentz violating results, if the 
$\mathbf{k}$ integral is performed on its natural, real domain. Now we give
details on these issues and later explain how Lorentz invariance and
analyticity are recovered.

The curve $\gamma $ is the boundary of the region $\mathcal{\tilde{A}}_{P}$.
It does not cross the real axis in $P$, but in the point $P^{\prime }$,
which has energy 
\begin{equation}
p^{0}=\sqrt{\frac{\mathbf{p}^{2}}{2}+\sqrt{\frac{(\mathbf{p}^{2})^{2}}{4}%
+4M^{4}}}\equiv E_{P^{\prime }}  \label{pp}
\end{equation}%
and satisfies $\sqrt{2}M<E_{P^{\prime }}<E_{P}$. Clearly, Lorentz invariance
is violated, because $P^{\prime }$ and $\gamma $ have no Lorentz invariant
meaning. This fact has been noticed by Nakanishi in ref. \cite{nakanishi}.
The intuitive reason is that, as shown in fig. \ref{wick}, the loop energy
is not everywhere real, so, if we want Lorentz invariance, the loop momentum
cannot be everywhere real. Said differently, if we want to restore \ Lorentz
invariance working at $\mathbf{p}\neq 0$, we must deform the $\mathbf{k}$
integration domain to include complex values, till the regions $\mathcal{%
\tilde{A}}_{i}$ are squeezed back to Lorentz invariant lines (i.e. solutions
of Lorentz invariant conditions), like those of fig. \ref{compl}. In
particular, the region $\mathcal{\tilde{A}}_{P}$ must be turned into the
half line $\mathcal{O}_{P}$ that corresponds to $p^{0}$ real located above
the LW threshold, i.e. $p^{0}\geqslant E_{P}$. During the deformation
process we can keep the deformed figure \ref{compl2} symmetric with respect
to the real axis. To achieve this goal, it is sufficient to separate the
contributions of the poles of each LW pair and deform the $\mathbf{k}$
integration domains in complex conjugate ways in the two cases.

Below we also show that when Lorentz invariance is violated (restored),
analyticity is also violated (restored).

\subsection{Lorentz invariance and analyticity above the LW\ threshold}

Now we study the amplitude in $\mathcal{O}_{P}$, its Lorentz invariance and
analyticity. It is convenient to separate the contributions of the physical
poles from the ones of the LW\ poles by writing the propagator (\ref{propa})
as%
\begin{equation}
iD_{0}(p^{2},m^{2},\epsilon )+iD_{\text{LW}}(p^{2},m^{2}),  \label{deLW}
\end{equation}%
where 
\begin{equation*}
D_{0}(p^{2},m^{2},\epsilon )=\frac{M^{4}}{M^{4}+m^{4}}\frac{1}{%
p^{2}-m^{2}+i\epsilon },\qquad D_{\text{LW}}(p^{2},m^{2})=-\frac{M^{4}}{%
M^{4}+m^{4}}\frac{p^{2}+m^{2}}{(p^{2})^{2}+M^{4}}.
\end{equation*}%
To simplify these expressions, we have replaced $m^{2}-i\epsilon $ with $%
m^{2}$ where allowed.

We just need to focus on the contribution 
\begin{equation}
\mathcal{J}_{\text{LW}}(p)=\int \frac{\mathrm{d}^{D}k}{(2\pi )^{D}}D_{\text{%
LW}}(k^{2},m_{1}^{2})D_{\text{LW}}((k-p)^{2},m_{2}^{2})  \label{JLW}
\end{equation}%
to the bubble loop integral $\mathcal{J}(p)$, because for $p$ real it is the
only one interested by the LW\ pinching. Every other contribution admits an
analytic Wick rotation.

We integrate on $k^{0}$ by means of the residue theorem, as usual, and
assume that $\mathbf{k}$ is integrated on its natural real domain. Then, the
function $\mathcal{J}_{\text{LW}}(p)$ is analytic and Lorentz invariant in
the main region $\mathcal{\tilde{A}}_{0}$, because the Wick rotation is
analytic there. It is neither analytic nor Lorentz invariant inside $%
\mathcal{\tilde{A}}_{P}$. Nevertheless, in the next section we prove that $%
\mathcal{J}_{\text{LW}}(p)$ is continuous everywhere if $\mathbf{p}\neq 0$.
We denote the function $\mathcal{J}_{\text{LW}}(p)$ restricted to $\mathcal{%
\tilde{A}}_{0}$ by $\mathcal{J}_{\text{LW}}^{0}(p)$ and the same function
restricted to $\mathcal{\tilde{A}}_{P}$ by $\mathcal{J}_{\text{LW}}^{P}(p)$.

When we deform the $\mathbf{k}$ integration domain, $\mathcal{J}_{\text{LW}%
}(p)$ changes into some new function $\mathcal{J}_{\text{LW}}^{\text{def}%
}(p) $, which depends on the deformation. Denote the deformed regions $%
\mathcal{\tilde{A}}_{0}$ and $\mathcal{\tilde{A}}_{P}$ by $\mathcal{\tilde{A}%
}_{0}^{\text{def}}$ and $\mathcal{\tilde{A}}_{P}^{\text{def}}$, respectively.

The function $\mathcal{J}_{\text{LW}}^{\text{def}}(p)$ is analytic in $%
\mathcal{\tilde{A}}_{0}^{\text{def}}$ and coincides with $\mathcal{J}_{\text{%
LW}}^{0}(p)$ in $\mathcal{\tilde{A}}_{0}\cap \mathcal{\tilde{A}}_{0}^{\text{%
def}}$. Moreover, as shown in the next section, it is continuous everywhere.
When the domain deformation is finalized, i.e. the surfaces of fig. \ref%
{compl2} are turned into the desired lines (in particular, $\mathcal{\tilde{A%
}}_{P}$ is squeezed onto $\mathcal{O}_{P}$), $\mathcal{J}_{\text{LW}}^{\text{%
def}}(p)$ gives the final outcome to be assigned to the integral (\ref{JLW})
in $\mathcal{O}_{P}$, which we denote by $\mathcal{J}_{\text{LW}}^{>}(p)$.

\begin{figure}[t]
\begin{center}
\includegraphics[width=10truecm]{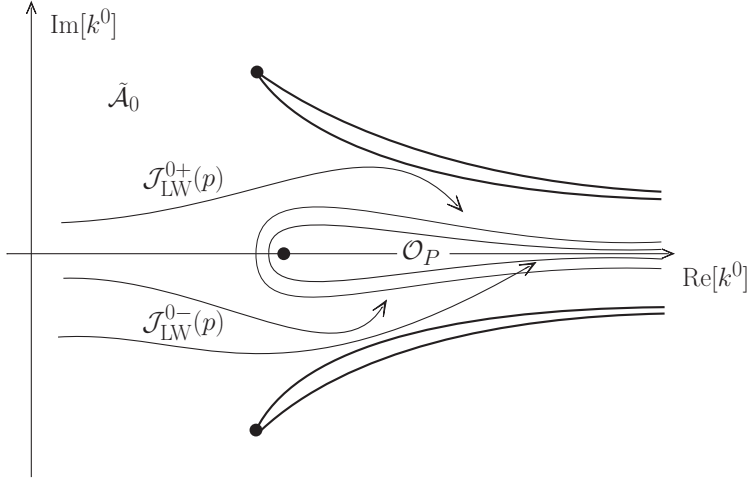}
\end{center}
\caption{Definitions of $\mathcal{J}_{\text{LW}}^{0+}(p)$ and $\mathcal{J}_{%
\text{LW}}^{0-}(p)$}
\label{complJW}
\end{figure}

We argue that%
\begin{equation}
\mathcal{J}_{\text{LW}}^{>}(p)=\frac{1}{2}\left[ \mathcal{J}_{\text{LW}%
}^{0+}(p)+\mathcal{J}_{\text{LW}}^{0-}(p)\right] ,  \label{jp}
\end{equation}%
where the functions $\mathcal{J}_{\text{LW}}^{0\pm }(p)$ are defined as
follows. Start from the function $\mathcal{J}_{\text{LW}}^{0}(p)$ in $%
\mathcal{\tilde{A}}_{0}$, which we know to be analytic. We can analytically
continue $\mathcal{J}_{\text{LW}}^{0}(p)$ to $\mathcal{O}_{P}$ either from
the half plane Im$[p^{0}]>0$ or from the half plane Im$[p^{0}]<0$, as shown
in fig. \ref{complJW}. These two possibilities give $\mathcal{J}_{\text{LW}%
}^{0+}(p)$ and $\mathcal{J}_{\text{LW}}^{0-}(p)$, respectively.

A number of arguments and checks, which we collect in the next section and
in section \ref{morecompl}, suggest that formula (\ref{jp}) is correct for
every diagram. Alternatively, we can take the right-hand side of (\ref{jp})
as the very \textit{definition} of $\mathcal{J}_{\text{LW}}^{>}(p)$,
bypassing the domain deformation described in the previous subsection.

The continuations that define $\mathcal{J}_{\text{LW}}^{0\pm }(p)$ in $%
\mathcal{O}_{P}$ can be stretched to neighborhoods of $\mathcal{O}_{P}$
above $P$, so both functions $\mathcal{J}_{\text{LW}}^{0\pm }(p)$ are
analytic in such neighborhoods. Moreover, they are Lorentz invariant,
because they are obtained from $\mathcal{J}_{\text{LW}}^{0}(p)$, which is
Lorentz invariant. Thus, formula (\ref{jp}) ensures that $\mathcal{J}_{\text{%
LW}}^{>}(p)$ is analytic and Lorentz invariant in a neighborhood of the real
axis above the LW threshold.

The function $\mathcal{J}_{\text{LW}}(p)$ is purely imaginary on the real
axis, because the integrand and the $\mathbf{k}$ integration domain are
real. Indeed, when we apply the residue theorem to integrate on $k^{0}$, we
pick pairs of complex conjugate poles and get an overall factor $i$. Thus, $%
\mathcal{J}_{\text{LW}}^{0}(p)=-[\mathcal{J}_{\text{LW}}^{0}(p^{\ast
})]^{\ast }$, which implies $\mathcal{J}_{\text{LW}}^{0-}(p)=-[\mathcal{J}_{%
\text{LW}}^{0+}(p^{\ast })]^{\ast }$. Since the contributions due to the
poles of each LW pair are interested by complex conjugate deformations of
the respective $\mathbf{k}$ integration domains, $\mathcal{J}_{\text{LW}}^{%
\text{def}}(p)$ obeys similar relations thoughout the deformation. Then, $%
\mathcal{J}_{\text{LW}}^{>}(p)$ is purely imaginary in $\mathcal{O}_{P}$ and
so is the right-hand side of (\ref{jp}). The amplitude $\mathcal{M}%
(p)=-i\lambda ^{2}\mathcal{J}(p)/2$ satisfies unitarity, because the LW
contributions do not affect its imaginary part on the real axis. More
details about unitarity can be found in ref. \cite{LWunitarity}.

\begin{figure}[t]
\begin{center}
\includegraphics[width=7truecm]{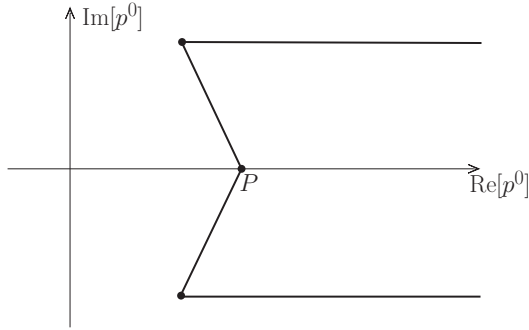}
\end{center}
\caption{Final analytic regions $\mathcal{A}_{i}$}
\label{compl3}
\end{figure}

It may be helpful to analytically continue the result from the mentioned
neighborhoods to larger regions. Focusing on the three regions that have
nontrivial intersections with the real axis, in the end we may get, for
example, a final figure like fig. \ref{compl3} plus its symmetrization with
respect to the imaginary axis. We see that the complex plane is divided into
the disjoint regions $\mathcal{A}_{0}$, $\mathcal{A}_{P}$ and $\mathcal{A}%
_{P}^{\prime }$, which are originated by the initial regions $\mathcal{%
\tilde{A}}_{0}$, $\mathcal{\tilde{A}}_{P}$ and $\mathcal{\tilde{A}}%
_{P}^{\prime }$ through the deformation process described previously.

Note that formula (\ref{jp}) allows us to find $\mathcal{J}_{\text{LW}%
}^{>}(p)$ without effectively going through the domain deformation process
(which is practically hard to implement): it is sufficient to decompose the
propagators as in formula (\ref{deLW}), isolate the contributions interested
by the LW pinching, analytically continue them from the main region $%
\mathcal{\tilde{A}}_{0}$ to $\mathcal{O}_{P}$ in the two possible ways and
finally average the results. As said, formula (\ref{jp}) could also be taken
as the definition of the function $\mathcal{J}_{\text{LW}}^{>}(p)$ in $%
\mathcal{O}_{P}$.

To summarize, the integral $\mathcal{J}(p)$ is ill defined at $\mathbf{p}=0$%
, but it can be worked out from $\mathbf{p}\neq 0$, without the need of 
\textit{ad hoc} prescriptions. We have derived the results in the case of
the bubble diagram, but the specificity of that diagram never really enters,
so\ we expect that the conclusions hold for every diagram. More comments on
this are contained in section \ref{morecompl}. An explicit check of the
result (\ref{jp}) is given in the next section [see the comments around
formula (\ref{ko})].

\section{Calculation around the LW pinching}

\setcounter{equation}{0}

\label{LWaround}

In this section we illustrate the calculations in the presence of the LW
pinching, prove the continuity of $\mathcal{J}_{\text{LW}}(p)$ and $\mathcal{%
J}_{\text{LW}}^{\text{def}}(p)$ and provide arguments and checks in favor of
formula (\ref{jp}). For definiteness, we assume to work in more than two
spacetime dimensions.

We focus on the pinching depicted in the left figure \ref{pinching2}. The $%
\mathbf{k}$ integral has potential singularities of the form $1/D_{0}$ and $%
1/D_{0}^{\ast }$, where%
\begin{equation}
D_{0}=p^{0}-\Omega _{M}(\mathbf{k})-\Omega _{M}^{\ast }(\mathbf{k-p}).
\label{D0}
\end{equation}%
The top pinching occurs for $D_{0}=0$, i.e.%
\begin{equation}
p^{0}=\sqrt{\mathbf{k}^{2}+iM^{2}}+\sqrt{(\mathbf{k}-\mathbf{p})^{2}-iM^{2}},
\label{ju}
\end{equation}%
while the bottom pinching occurs for $D_{0}^{\ast }=0$. The conditions are
complex for $\mathbf{p}\neq 0$, so they split into two real conditions.

We want to study $\mathcal{J}(p)$ above the LW threshold, so we take a real $%
p^{0}>\sqrt{\mathbf{p}^{2}+2M^{2}}$. With a real loop space momentum $%
\mathbf{k}$, the solution of (\ref{ju}) is a circle, equal to the
intersection between a sphere and a plane, given by%
\begin{equation}
\mathbf{k}^{2}=\frac{(p^{0})^{4}-4M^{4}}{4(p^{0})^{2}},\qquad \mathbf{p}%
^{2}=2\mathbf{p}\cdot \mathbf{k}.  \label{solution}
\end{equation}%
If the external energy $p^{0}$ is complex, the analysis becomes more
involved, but for our purposes it is sufficient to focus on the values of $%
p^{0}$ that are close to the real axis. This can be achieved by making the
substitution $p^{0}\rightarrow p^{0}\mathrm{e}^{i\varphi }$, with $\varphi $
small, after which we can keep $p^{0}$ real. The denominator becomes%
\begin{equation*}
D_{\varphi }=p^{0}\mathrm{e}^{i\varphi }-\Omega _{M}(\mathbf{k})-\Omega
_{M}^{\ast }(\mathbf{k-p}).
\end{equation*}%
To simplify the formulas, we expand $D_{\varphi }$ around the solution (\ref%
{solution}) by means of the change of variables 
\begin{equation}
k_{s}=\frac{\sigma _{-}}{2p^{0}}+\tau \frac{\sigma _{+}^{2}}{2\sigma
_{-}(p^{0})^{2}}+\eta \frac{p_{s}\sigma _{+}^{2}}{4\sigma _{-}M^{2}},\qquad
u=\frac{p_{s}}{2k_{s}}+\eta \frac{\sigma _{+}^{2}}{2\sigma _{-}M^{2}},
\label{para}
\end{equation}%
where $\sigma _{\pm }\equiv \sqrt{(p^{0})^{4}\pm 4M^{4}}$, $p_{s}\equiv |%
\mathbf{p}|$ and $u=\cos \theta $, $\theta $ being the angle between the
vectors $\mathbf{p}$ and $\mathbf{k}$. The fluctuations around the solutions
(\ref{solution}) are parametrized by $\tau $ and $\eta $. The integrand of $%
\mathcal{J}(p)$ is proportional to 
\begin{equation}
\frac{\mathrm{d}^{D-1}\mathbf{k}}{D_{\varphi }}\rightarrow -\frac{2\pi
^{(D-2)/2}}{\Gamma \left( \frac{D}{2}-1\right) }\frac{%
k_{s}^{D-2}(1-u^{2})^{(D-4)/2}\mathrm{d}k_{s}\mathrm{d}u}{\tau
-i(p^{0}\varphi +p_{s}\eta )},  \label{te}
\end{equation}%
where the arrow means that we have integrated on all the angles besides $%
\theta $. We have also expanded the denominator to the first order in $%
\varphi $, $\tau $ and $\eta .$

We see that as long as either $\varphi $ or $p_{s}$ are different from zero,
the potential singularity at $D_{\varphi }=0$ is integrable. In particular,
if we keep $p_{s}\neq 0$ and reach $\varphi =0$, we obtain%
\begin{equation}
\frac{\mathrm{d}^{D-1}\mathbf{k}}{D_{0}}\rightarrow -\frac{2\pi ^{(D-2)/2}}{%
\Gamma \left( \frac{D}{2}-1\right) }\frac{\sigma _{+}^{4}\sigma _{-}^{D-4}}{%
(2p^{0})^{D}M^{2}}(1-u^{2})^{(D-4)/2}\frac{\mathrm{d}\tau \mathrm{d}\eta }{%
\tau -ip_{s}\eta }.  \label{prescr}
\end{equation}%
It is interesting to study the limit $p_{s}\rightarrow 0$ of this
expression, which gives%
\begin{equation}
-\frac{4\pi ^{(D-2)/2}}{\Gamma \left( \frac{D}{2}-1\right) }\frac{\sigma
_{+}^{2}\sigma _{-}^{D-3}}{(2p^{0})^{D}}(1-u^{2})^{(D-4)/2}\mathrm{d}\tau 
\mathrm{d}u\left[ \mathcal{P}\left( \frac{1}{\tau }\right) +i\pi \mathrm{sgn}%
(u)\delta (\tau )\right] ,  \label{ter}
\end{equation}%
where $\mathcal{P}$ denotes the principal value and sgn is the sign
function. We learn that in this case $p_{s}$ provides the prescription for
handling the integral. Note that at $p_{s}=0$ no $u$ dependence survives in
the integrand, besides the sign function of formula (\ref{ter}) and the
factor $(1-u^{2})^{(D-4)/2}$ coming from the integration measure. If we
perform the simple $u$ integration, we finally get%
\begin{equation}
-\frac{4\pi ^{(D-1)/2}}{\Gamma \left( \frac{D-1}{2}\right) }\frac{\sigma
_{+}^{2}\sigma _{-}^{D-3}}{(2p^{0})^{D}}\mathrm{d}\tau \mathcal{P}\left( 
\frac{1}{\tau }\right) .  \label{terra}
\end{equation}%
Also note that in three and higher dimensions there is no singularity for $%
\sigma _{-}\rightarrow 0^{+}$.

Formula (\ref{prescr}), applied to $\mathcal{J}_{\text{LW}}(p)$ at $%
p_{s}\neq 0$, shows that\ $\mathcal{J}_{\text{LW}}(p)$ is
continuous everywhere on the complex $p^{0}$ plane, as anticipated in the
previous section. We have also checked the continuity of $\mathcal{J}_{\text{%
LW}}(p)$ numerically, by means of a computer program.

If we use formula (\ref{terra}) in $\mathcal{J}_{\text{LW}}(p)$, we can work
out the function $\mathcal{J}_{\text{LW}}^{>}(p)$ for $p_{s}\rightarrow 0$.
Indeed, having set $\varphi =0$ we have placed ourselves in $\mathcal{O}%
_{P}\subset \mathcal{\tilde{A}}_{P}$. This allows us to evaluate the
integral $\mathcal{J}_{\text{LW}}(p)$ there at $p_{s}\neq 0$. Then, the
limit $p_{s}\rightarrow 0$ squeezes the region $\mathcal{\tilde{A}}_{P}$
onto $\mathcal{O}_{P}$ and so gives $\mathcal{J}_{\text{LW}}^{>}(p)$. Here,
it is unnecessary to actually perform the domain deformation, because the
limit $p_{s}\rightarrow 0$ provides an equivalent effect.

Nowe we can check formula (\ref{jp}), proceeding as follows. We study the
singularity $1/D_{\varphi }$ again, but first set $p_{s}=0$ at nonzero $%
\varphi $ and then send $\varphi $ to zero. By formula (\ref{te}), the
denominator $\tau -ip_{s}\eta $ of (\ref{prescr}) is replaced by $\tau
-ip^{0}\varphi $, so, after integrating on $u$, we find%
\begin{equation}
-\frac{4\pi ^{(D-1)/2}}{\Gamma \left( \frac{D-1}{2}\right) }\frac{\sigma
_{+}^{2}\sigma _{-}^{D-3}}{(2p^{0})^{D}}\frac{\mathrm{d}\tau }{\tau
-ip^{0}\varphi }\underset{\varphi \rightarrow 0^{\pm }}{\longrightarrow }-%
\frac{4\pi ^{(D-1)/2}}{\Gamma \left( \frac{D-1}{2}\right) }\frac{\sigma
_{+}^{2}\sigma _{-}^{D-3}}{(2p^{0})^{D}}\mathrm{d}\tau \left[ \mathcal{P}%
\left( \frac{1}{\tau }\right) \pm i\pi \delta (\tau )\right] .  \label{ko}
\end{equation}%
These expressions are also regular, but do not coincide with (\ref{terra}).

Observe that the result (\ref{ko}) is obtained by first squeezing the region 
$\mathcal{\tilde{A}}_{P}$ onto $\mathcal{O}_{P}$ (which is a consequence of
letting $p_{s}\rightarrow 0$ first) and then approaching the real axis from $%
\mathrm{Im}[p^{0}]>0$ ($\varphi \rightarrow 0^{+}$) or $\mathrm{Im}[p^{0}]<0$
($\varphi \rightarrow 0^{-}$). The two cases give $\mathcal{J}_{\text{LW}%
}^{0+}(p)$ and $\mathcal{J}_{\text{LW}}^{0-}(p)$, respectively. If we
average the two results (\ref{ko}), we obtain (\ref{terra}), in agreement
with formula (\ref{jp}).

We expect that the key results just found continue to hold through the
domain deformation that defines the amplitude in $\mathcal{O}_{P}$ at $%
\mathbf{p}\neq 0$. For example, the basic reason why $\mathcal{J}_{\text{LW}%
}(p)$ is continuous everywhere is that the denominator $D_{\varphi }$ is
complex, which makes the singularity integrable. However, the denominator
remains complex during the domain deformation, so $\mathcal{J}_{\text{LW}}^{%
\text{def}}(p)$ is also continuous. Moreover, the check of formula (\ref{ju}%
) provided above, which works at $p_{s}\rightarrow 0$, captures the
essential features that also apply at $\mathbf{p}\neq 0$, when the domain
deformation is taken into account. Indeed, assume that the deformed region $%
\mathcal{\tilde{A}}_{P}^{\text{def}}$ is a thin strip around $\mathcal{O}%
_{P} $. Let $\tilde{p}_{s}$ denote the length of the short edge of the
strip, so that the domain deformation is finalized ($\mathcal{\tilde{A}}%
_{P}^{\text{def}}\rightarrow \mathcal{O}_{P}$) when $\tilde{p}%
_{s}\rightarrow 0$. On general grounds, the potential singularity of the
integral is always expected to be of the form%
\begin{equation}
\sim \frac{\mathrm{d}\tau \mathrm{d}\eta }{\tau -i(p^{0}\varphi +\tilde{p}%
_{s}\eta )},  \label{stima}
\end{equation}%
where the external momentum is still written as $p^{0}\mathrm{e}^{i\varphi }$%
, with $p^{0}$ real and $\varphi $ small, while $\tau $ and $\eta $ are two
real variables that parametrize the fluctuations around the singular point
at $\varphi =0$ ($\tau $ being parallel to the long edge of the strip and $%
\eta $ being parallel to the short edge). Repeating the arguments above with
the help of (\ref{stima}), we still get formula (\ref{jp}).

The results just derived and formula (\ref{jp}) are expected to apply to the
LW pinching of any diagram, because they are not tied to the peculiarities
to the bubble diagram. See section \ref{morecompl} for more details on this.

\subsection{Comparison with the CLOP and other prescriptions}

We have seen that the theory is intrinsically equipped with the right recipe
to handle the LW pinching. This means that any artificial prescription can
potentially lead to wrong results. Now we classify the whole set of unitary
prescriptions, which includes the CLOP one, and compare them with the
results predicted by the formulation of this paper. For definiteness, we
work in four dimensions.

Consider the integrand of the loop integral (\ref{bubd}) at $\mathbf{p}=0$.
We begin with the top pinching that appears in the left figure \ref%
{pinching2}, which is due to the poles (\ref{polpot}). By means of the
expansion%
\begin{equation}
k_{s}=\frac{\sigma _{-}}{2p^{0}}+\tau \frac{Mp^{0}}{\sigma _{-}},  \label{ks}
\end{equation}%
we see that the integrand of $\mathcal{J}(p)$ behaves as%
\begin{equation}
\frac{i}{(8\pi )^{2}}\frac{\sigma _{-}}{(p^{0})^{2}}\frac{M^{4}}{%
(M^{2}+im_{1}^{2})(M^{2}-im_{2}^{2})}\frac{\mathrm{d}\tau }{\tau }
\label{pole}
\end{equation}%
around the singularity $\tau =0$.

We know that the formulation of this paper removes the singularity because,
working at nonvanishing $\mathbf{p}$ and letting $\mathbf{p}$ tend to zero
afterwards, (\ref{pole}) is replaced by%
\begin{equation}
\frac{i}{(8\pi )^{2}}\frac{\sigma _{-}}{(p^{0})^{2}}\frac{M^{4}}{%
(M^{2}+im_{1}^{2})(M^{2}-im_{2}^{2})}\mathcal{P}\left( \frac{1}{\tau }%
\right) \mathrm{d}\tau .  \label{ours}
\end{equation}%
More generally, we may have%
\begin{equation}
\frac{i}{(8\pi )^{2}}\frac{\sigma _{-}}{(p^{0})^{2}}\frac{M^{4}}{%
(M^{2}+im_{1}^{2})(M^{2}-im_{2}^{2})}\left[ \mathcal{P}\left( \frac{1}{\tau }%
\right) +ia\delta (\tau )\right] \mathrm{d}\tau ,  \label{pola}
\end{equation}%
where $a$ is an arbitrary real constant.

The LW\ poles come in conjugate pairs, so the pinching just considered is
accompanied by the complex conjugate one, which occurs when\ the residue
calculated in $k^{0}=p^{0}-\Omega _{M}(\mathbf{k})$ hits the LW pole located
in $k^{0}=\Omega _{M}^{\ast }(\mathbf{k})$. The contribution is minus the
complex conjugate of (\ref{pola}), because the $i$ factor that accompanies
the residue does not get conjugated. The total gives%
\begin{equation*}
\frac{2i}{(8\pi )^{2}}\frac{\sigma _{-}}{(p^{0})^{2}}\frac{M^{4}}{%
(M^{4}+m_{1}^{4})(M^{4}+m_{2}^{4})}\left[ (M^{4}+m_{1}^{2}m_{2}^{2})\mathcal{%
P}\left( \frac{1}{\tau }\right) +aM^{2}(m_{1}^{2}-m_{2}^{2})\delta (\tau )%
\right] \mathrm{d}\tau .
\end{equation*}%
Again, the contribution to $\mathcal{J}(p)$ is regular and purely imaginary.
In particular, it does not affect the imaginary part of the amplitude $%
\mathcal{M}(p)=-i\lambda ^{2}\mathcal{J}(p)/2$. This result proves that the
prescription (\ref{pola}) is consistent with perturbative unitarity for
arbitrary $a$. However, the loop integral $\mathcal{J}(p)$ does depend on $a$%
, at least when the two physical masses are different. This proves that no
prescription with nonvanishing $a$ is consistent with our formulation, which
predicts $a=0$.

The CLOP prescription is ambiguous and gives $a=\pm \pi $. This result can
be proved by replacing the LW scale $M$ with $M^{\prime }=M+\delta $ in the
second propagator of (\ref{bubd}). Modifying the expansion (\ref{ks}) into%
\begin{equation*}
k_{s}=\frac{\sigma _{-}}{2p^{0}}+\tau \frac{Mp^{0}}{\sigma _{-}}-2\delta 
\frac{M^{3}}{p^{0}\sigma _{-}},
\end{equation*}%
the integrand $\mathcal{J}(p)$ behaves as%
\begin{equation}
\frac{i}{(8\pi )^{2}}\frac{\sigma _{-}}{(p^{0})^{2}}\frac{iM^{4}}{%
(M^{2}+im_{1}^{2})(M^{2}-im_{2}^{2})}\frac{\mathrm{d}\tau }{\tau -i\delta },
\label{cloppa}
\end{equation}%
around the top pinching of the left figure \ref{pinching2}. This formula is
equivalent to (\ref{pola}) with $a=\pi $sgn$(M^{\prime }-M)$.

The result is ambiguous, because it depends on whether $\delta $ is chosen
to be positive or negative and there is no way to decide whether $M^{\prime
} $ must be smaller than $M$ or the contrary. In the next section we plot
the ambiguity numerically.

Before this result, ambiguities due to the CLOP\ prescription were expected
only in more complicated diagrams \cite{CLOP}. It was understood that maybe
it was possible to resolve them by means of further prescriptions. The
ambiguity we have just found is present already at one loop and in one of
the simplest Feynman diagrams. However, it occurs only when $m_{1}\neq m_{2}$%
, which explains why it was not noticed before. For example, the results of
ref. \cite{grinstein} are correct, since they are made in the case $%
m_{1}=m_{2}=m$, where the CLOP\ prescription gives the same result as our
formulation.

If we want to make the new ambiguity disappear, we can supplement the CLOP
prescription by an average over the two possibilities $a=\pm \pi $, which
effectively gives $a=0$ and agrees with our result (\ref{ours}). This makes
the amended prescription even more artificial than the original CLOP
approach and there is still no guarantee that analogous way outs can be
found in more complicated situations. For these reasons, we think that 
\textit{ad hoc} approaches like the CLOP one should be dropped in favor of
the new formulation of this paper, which does not have such problems.

\section{ Complete bubble diagram}

\label{bubblecomplete} \setcounter{equation}{0}

In this section, we complete the calculation of the bubble diagram. The main
goal is to describe what happens around the LW threshold. Since the
threshold associated with the physical poles is not the main focus of the
calculation, we avoid the superposition between the physical threshold and
the LW one by assuming that the masses $m_{1}$ and $m_{2}$ are sufficiently
large. For concreteness, we take $m_{1},m_{2}\geqslant 3M$.

Another simplifying choice is to make the calculation at $\mathbf{p}=0$ and
resolve the singularity with the help of formula (\ref{terra}). We know that
this procedure is justified by starting from nonvanishing $\mathbf{p}$,
where the LW pinching is properly handled, and taking the limit $\mathbf{p}%
\rightarrow 0$ afterwards.

\begin{figure}[t]
\begin{center}
\includegraphics[width=8truecm]{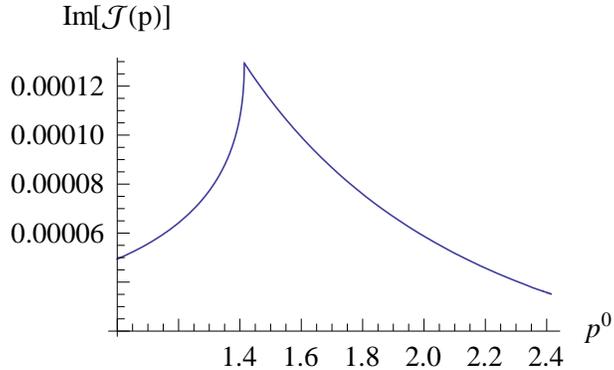}
\end{center}
\caption{Plot of $\mathrm{Im}[\mathcal{J}(p)]$ around the LW pinching}
\label{Bubble}
\end{figure}

Setting $M=1$ and $m_{1}=m_{2}=3$, the imaginary part of $\mathcal{J}(p)$ as
a function of a real $p^{0}$ has the behavior of fig. \ref{Bubble}. The real
part vanishes in the range shown, in agreement with unitarity. We see that
the imaginary part is well defined and continuous, but not analytic. The
nonanalyticity that is visible at $p^{2}=2M^{2}$ is the remnant of the LW\
pinching. If in nature some physical processes are described by a LW theory,
the LW scale $M$ is the key physical quantity signaling the new physics. A
shape like the one of fig. \ref{Bubble} may be helpful to determine the
magnitude of $M$ experimentally.

The formulation of the theory by nonanalytically Wick rotating its Euclidean
version gives an unambiguous answer and does not need \textit{ad hoc}
prescriptions. The CLOP prescription gives the same result, in the case just
considered. As explained in the previous section, we can appreciate the
intrinsic ambiguity of the CLOP\ prescription and the differences with the
predictions of our formulation by studying the bubble diagram with unequal
masses. For example, we compare the case $m_{1}=3$, $m_{2}=5$ to the case $%
m_{1}=m_{2}=4$.

Using the CLOP prescription, we take $M=1$ in the first propagator of
formula (\ref{bubd}) and $M=1+\delta $ in the second propagator, working at $%
\mathbf{p}=0$. Then we integrate $\mathcal{J}(p)$ numerically for smaller
and smaller values of $|\delta |$, till, say, $|\delta |=10^{-3}$. We study
both $\delta =-10^{-3}$ and $\delta =10^{-3}$.

On the other hand, following the formulation proposed here, we set $M=1$ in
both propagators, but keep $p_{s}=|\mathbf{p}|$ different from zero. Then,
we integrate numerically for smaller and smaller values of $p_{s}$ till $%
p_{s}=10^{-3}$.

\begin{figure}[t]
\begin{center}
\includegraphics[width=7.5truecm]{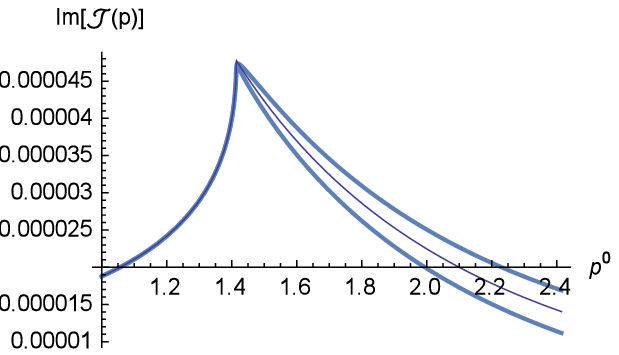}\quad %
\includegraphics[width=7.5truecm]{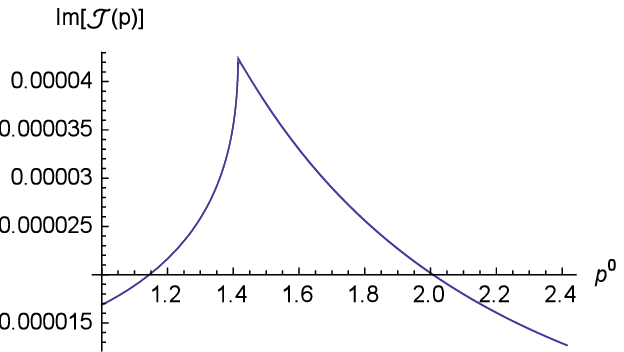}
\end{center}
\caption{Comparison between our formulation and the CLOP prescription}
\label{noCLOP}
\end{figure}

Collecting the results of these calculations, the imaginary part of $%
\mathcal{J}(p)$ gives the plots of fig. \ref{noCLOP}, while the real part
still vanishes. The first plot refers to the case $m_{1}=3$, $m_{2}=5$,
while the second plot refers to the case $m_{1}=m_{2}=4$. Let us describe
the first plot in detail. Below the LW\ threshold, the graph is unique,
which means that our formulation and the CLOP\ prescription give the same
result. Above the LW threshold, we see three graphs. The middle graph is the
one predicted by our formulation, while the upper and lower graphs are those
predicted by the CLOP prescription, with $\delta =-10^{-3}$ and $\delta
=10^{-3}$, respectively.

Although the match is very precise in the equal mass case (second figure),
there is a remarkable discrepancy above the LW threshold in the unequal mass
case. These results confirm that the CLOP prescription gives two different
results depending on whether $M^{\prime }>M$ or $M^{\prime }<M$. The average
of the two CLOP\ graphs coincides with the graph predicted by our
formulation.

If we really want to retrieve our result from a procedure where the
propagators of formula (\ref{bubd}) have two different LW scales $M$ and $%
M^{\prime }$, as in the CLOP prescription, we actually can, but in that case
the CLOP\ prescription becomes redundant. Instead of setting $p_{s}=0$ and
then letting $M^{\prime }$ tend to $M$, we must start from $p_{s}\neq 0$,
let $M^{\prime }$ approach $M$ while $p_{s}\neq 0$, work in a suitable
region $\mathcal{\tilde{A}}_{>}$, perform the domain deformation\ and only
at the end, if we want, let $p_{s}$ tend to zero.

\begin{figure}[b]
\begin{center}
\includegraphics[width=7truecm]{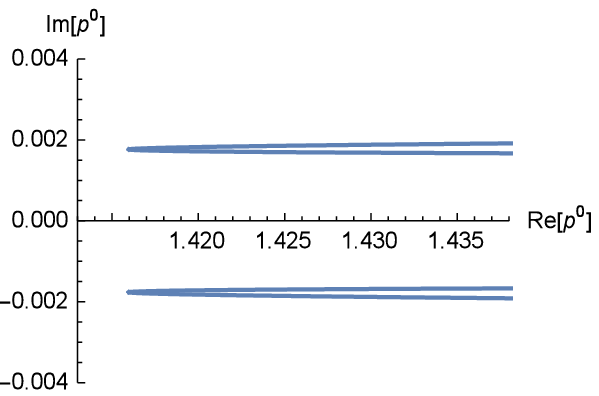}\hskip1truecm%
\includegraphics[width=7truecm]{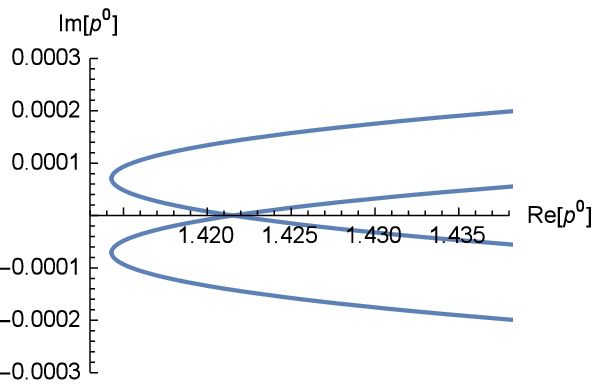}
\end{center}
\caption{Areas of LW pinching when $M^\prime\neq M$}
\label{CLOPcompl}
\end{figure}

In more detail, the region $\mathcal{\tilde{A}}_{P}$ contained in the curve $%
\gamma $ of fig. \ref{compl2} splits into two regions $\mathcal{\tilde{A}}%
_{P}^{+}$ and $\mathcal{\tilde{A}}_{P}^{-}$, when $M^{\prime }=1+\delta $ is
sufficiently different from $M$ (or $p_{s}$ is sufficiently large). We show
the new regions in fig. \ref{CLOPcompl}, where we have taken $p_{s}=10^{-3}$
and $M=1$. In the left picture $\delta =5\cdot 10^{-3}$, while in the right
picture $\delta =10^{-4}$.

When $\delta $ is sufficiently large, the real axis has no intersection with 
$\mathcal{\tilde{A}}_{P}^{+}$ and $\mathcal{\tilde{A}}_{P}^{-}$, but when $%
\delta $ becomes smaller, the region $\mathcal{\tilde{A}}_{>}\equiv \mathcal{%
\tilde{A}}_{P}^{+}\cap \mathcal{\tilde{A}}_{P}^{-}$ is nonempty. What the
CLOP\ prescription requires is to cover the entire real axis by analytic
continuation from below (i.e. from the region that contains the imaginary
axis) and let $\delta $ tend to zero at the end. What our formulation
requires, instead, is to reach the portion of the real axis that is located
above the LW threshold $P$ by working in $\mathcal{\tilde{A}}_{>}$, perform
the domain deformation that squeezes $\mathcal{\tilde{A}}_{>}$ onto the real
axis, let $\delta $ tend to zero and finally analytically continue the
result to reach $P$ from above. This is the crucial difference between the
two formulations, which explains the discrepancy shown in fig. \ref{noCLOP}.

\section{More complicated diagrams}

\setcounter{equation}{0}

\label{morecompl}

In this section, we explain how the arguments of the previous sections can
be extended to more complicated diagrams. One-loop diagrams have a unique
loop momentum, while the independent external momenta can be arbitrarily
many. The pure LW pinchings are similar to the ones of the bubble diagram.
They occur between the right LW\ poles of any propagator and the left LW
poles of any other propagator, as described by figure \ref{WickBub2}. The
mixed LW pinching cannot occur for real external momenta.

At higher loops the pinching is also analogous to the one we are accustomed
to in common theories. There, if the propagators of the internal legs of the
diagram have masses $m_{i}$, the pinchings lead to thresholds of the form $%
p^{2}=(m_{i_{1}}+m_{i_{2}}+m_{i_{3}}+\cdots )^{2}$, where $p$ is a sum of
incoming momenta. In the case of the LW\ pinching, the formulas that give
the thresholds are basically the same, with the difference that some masses $%
m_{i}$ are replaced by the complex masses $M_{\pm }=(1\pm i)M/\sqrt{2}$
associated with the LW\ scales. The pinching conditions are always of the
form $p^{0}=$ positive sum of (possibly complex) frequencies and the
thresholds are%
\begin{equation*}
p^{2}=\left[ (n_{+}+n_{-})\frac{M}{\sqrt{2}}+i(n_{+}-n_{-})\frac{M}{\sqrt{2}}%
+m_{i_{1}}+m_{i_{2}}+m_{i_{3}}+\cdots \right] ^{2},
\end{equation*}%
where the integers $n_{+}$ and $n_{-}$ count how many times the masses $%
M_{+} $ and $M_{-}$ appear, respectively. The number of thresholds grows
with the number of loops and so does the number of disjoint regions $%
\mathcal{\tilde{A}}_{i}$ and $\mathcal{A}_{i}$. The thresholds that are
relevant to the calculations of the physical amplitudes are those that are
located on the real axis, which are 
\begin{equation*}
p^{2}=(\sqrt{2}nM+m_{i_{1}}+m_{i_{2}}+m_{i_{3}}+\cdots )^{2},
\end{equation*}%
where $n=n_{+}=n_{-}$.

We expect that the arguments of sections \ref{LWpin} and \ref{LWaround} for
the calculation around the LW\ pinching work in any diagram. In a generic
Lorentz frame, the regions $\mathcal{\tilde{A}}_{i}$ are enlarged. Lorentz
invariance is violated and the integration domain on the loop space momenta
must be deformed to recover it. Consider the behavior of a loop integral
around some LW pinching. During the domain deformation, the deformed surface 
$\mathcal{\tilde{A}}_{P}^{\text{def}}$ eventually becomes a thin strip
almost squeezed onto the real axis above the LW threshold $P$. If $\tau $
denotes a coordinate for the long edge of the strip and $\eta $ a coordinate
for the perpendicular edge, while $\tilde{p}_{s}$ measures the length of the
short edge, the denominator of (\ref{stima}) appears to capture the most
general behavior we can meet ($p^{0}$ being replaced by $p^{0}\mathrm{e}%
^{i\varphi }$, $\varphi $ small). Then, formula (\ref{jp}) is also expected
to hold, as well as Lorentz invariance and analyticity above the LW
thresholds.

The analytic regions $\mathcal{A}_{i}$ are determined as follows. Working in
a generic Lorentz frame, we find the regions $\mathcal{\tilde{A}}_{i}$ by
integrating on the natural, real domains of the loop space momenta [see fig. %
\ref{compl2}]. Decomposing the propagators as in formula (\ref{deLW}), we
isolate the contributions $\mathcal{J}_{\text{LW}}(p)$ interested by the LW
pinching. For each of them, we compute $\mathcal{J}_{\text{LW}}(p)$ in the
main region $\mathcal{\tilde{A}}_{0}$, which is the one that contains the
imaginary axis. Then we analytically continue the result \textquotedblleft
from below\textquotedblright , which means from smaller to larger values of
the squared external momentum $p^{2}$, till we reach a LW threshold $P$. We
proceed with the continuation above $P$, but here we find two different
functions, $\mathcal{J}_{\text{LW}}^{0+}(p)$ and $\mathcal{J}_{\text{LW}%
}^{0-}(p)$, depending on whether we continue from the half plane with $%
\mathrm{Im}[p^{0}]>0$ or the one with $\mathrm{Im}[p^{0}]<0$. By formula (%
\ref{jp}), the final outcome $\mathcal{J}_{\text{LW}}^{>}(p)$ to be assigned
to $\mathcal{J}_{\text{LW}}(p)$ above $P$, is the average of $\mathcal{J}_{%
\text{LW}}^{0+}(p)$ and $\mathcal{J}_{\text{LW}}^{0-}(p)$. It is Lorentz
invariant and can be analytically extended from the real axis to a region $%
\mathcal{A}_{P}$ whose boundary intersects the real axis only in $P\ $ (see
fig. \ref{compl3}). Following these directions, we obtain $\mathcal{J}_{%
\text{LW}}^{>}(p)$ without having to go through the domain deformation
process described in section \ref{LWpin}. The procedure must be applied to
every LW\ threshold $P$ and can be generalized to regions that are placed
above more LW thresholds at the same time. The final main region $\mathcal{A}%
_{0}$ is the complement of $\cup _{P}\mathcal{A}_{P}$.

\section{Conclusions}

\setcounter{equation}{0}

The Lee-Wick models are higher-derivative theories that are claimed to
reconcile renormalizability and unitarity in a very nontrivial way. However,
several aspects of their formulation remained unclear. In this paper, we
have provided a new formulation of the models that overcomes the major
difficulties, by defining them as nonanalytically Wick rotated Euclidean
theories. Working in a generic Lorentz frame, the models are intrinsically
equipped with the right recipe to treat the pinchings of the Lee-Wick poles,
with no need of external \textit{ad hoc} prescriptions. The complex energy
plane is divided into disconnected analytic regions, which are related to
one another by a well defined, albeit nonanalytic procedure.

The nonanalytic behaviors of the amplitudes may have interesting
phenomenological consequences, which may facilitate the measurements of some
key physical constants of the theories, such as the scales associated with
the higher-derivative terms.

\vskip12truept \noindent {\large \textbf{Acknowledgments}}

\vskip 12truept

We are grateful to U.G. Aglietti and L. Bracci for useful discussions.

\end{document}